\documentclass{ar-1col}
\usepackage[numbers]{natbib}
\usepackage{amssymb}
\usepackage{amsmath}
\usepackage{amsfonts}

\begin{document}

\setcounter{secnumdepth}{4}

\jname{Xxxx. Xxx. Xxx. Xxx.}
\jvol{AA}
\jyear{YYYY}
\doi{10.1146/((please add article doi))}


\markboth{Greene et al.}{The Strange Metal State of the Electron-Doped Cuprates}

\title{The Strange Metal State of the Electron-Doped Cuprates}

\author{Richard L. Greene,$^1$ Pampa R. Mandal,$^1$ Nicholas R. Poniatowski,$^1$ and Tarapada Sarkar$^1$
\affil{$^1$Center for Nanophysics and Advanced Materials and Department of Physics, University of Maryland, College Park, Maryland 20742}
}

\begin{abstract}
An understanding of the high-temperature copper oxide (cuprate) superconductors has eluded the physics community for over thirty years, and represents one of the greatest unsolved problems in condensed matter physics. Particularly enigmatic is the normal state from which superconductivity emerges, so much so that this phase has been dubbed a ``strange metal.'' In this article, we will review recent research into this strange metallic state as realized in the electron-doped cuprates with a focus on their transport properties. The electron-doped compounds differ in several ways from their more thoroughly studied hole-doped counterparts, and understanding these asymmetries of the phase diagram may prove crucial to developing a final theory of the cuprates. Most of the experimental results discussed in this review have yet to be explained and remain an outstanding challenge for theory.
\end{abstract}

\begin{keywords}
cuprates, quantum critical point, strange metal, scale-invariant transport
\end{keywords}
\maketitle


\section{INTRODUCTION}

The discovery of high-temperature superconductivity in the tetragonal ``214'' copper oxide $\mathrm{La}_{2-x} \mathrm{Ba}_x \mathrm{CuO}_4$ in 1986 \cite{bednorz86} was a seminal event in the history of condensed matter physics. However, after thirty years and over 100,000 publications, the mechanism that gives rise to superconductivity (SC) in the cuprates, and even the physics of their normal state, remains a mystery. The undoped parent compounds of these materials are known to be antiferromagnetic (AF) Mott insulators, and it is widely accepted that strong electron correlations play a central role in both the SC and normal states \cite{keimer16,rlgrev,scalapino12, lee06}.

\begin{marginnote}[]
\entry{SC}{Superconductivity}
\entry{AF}{Antiferromagnetism}
\entry{Cuprates}{A family of materials comprised of layered copper oxygen planes, which exhibit high-temperature SC when doped with electrons or holes}
\end{marginnote}

As the $\mathrm{CuO}_2$ planes are doped with charge carriers the AF phase is suppressed and SC emerges, as shown in the schematic phase diagram for n-type cuprates in Fig. 1. Despite the qualitative differences between hole- and electron-doped materials –- namely the disparate sizes of the AF and SC phases and the presence of a ``pseudogap'' on the hole-doped side \cite{statt99, norman05} -- the cause of the SC and the nature of the strange metallic normal state is most likely the same for both families. This is simply because both properties are driven by electron interactions within the $\mathrm{CuO}_2$ plane, which is a universal feature of all cuprates owing to their anisotropic (2D) structure that drastically weakens out-of-plane (interlayer) coupling.  

\begin{marginnote}[]
\entry{Fermi Liquid}{A metal well-described by Landau's Fermi liquid theory, which predicts a $T^2$ dependence of the resistivity and a $H^2$ dependence of the magnetoresistance}
\end{marginnote}

In this review, we will focus on insights gleaned from the study of electron-doped (n-type) cuprates, and draw connections to the hole-doped (p-type) compounds where appropriate. The n-type cuprates have several attractive features which simplify their experimental study (and perhaps their theoretical understanding as well), in particular the absence of a pseudogap and a small upper critical field ($H_{c2} < 10$T) that enables measurement of the normal state down to mK temperatures. We will further narrow our focus to the normal state properties of these materials, with special emphasis on their unconventional (i.e. strange metallic) transport properties as a function of temperature, magnetic field, and doping. We will assume that there is some correspondence between the emergence of high-temperature SC from this strange metallic phase and the emergence of electron-phonon SC from a conventional Fermi liquid, and will not discuss the SC state itself in any detail, as it is already adequately reviewed in the literature \cite{rlgrev}.

\begin{figure}
\centering{\includegraphics[width=70mm]{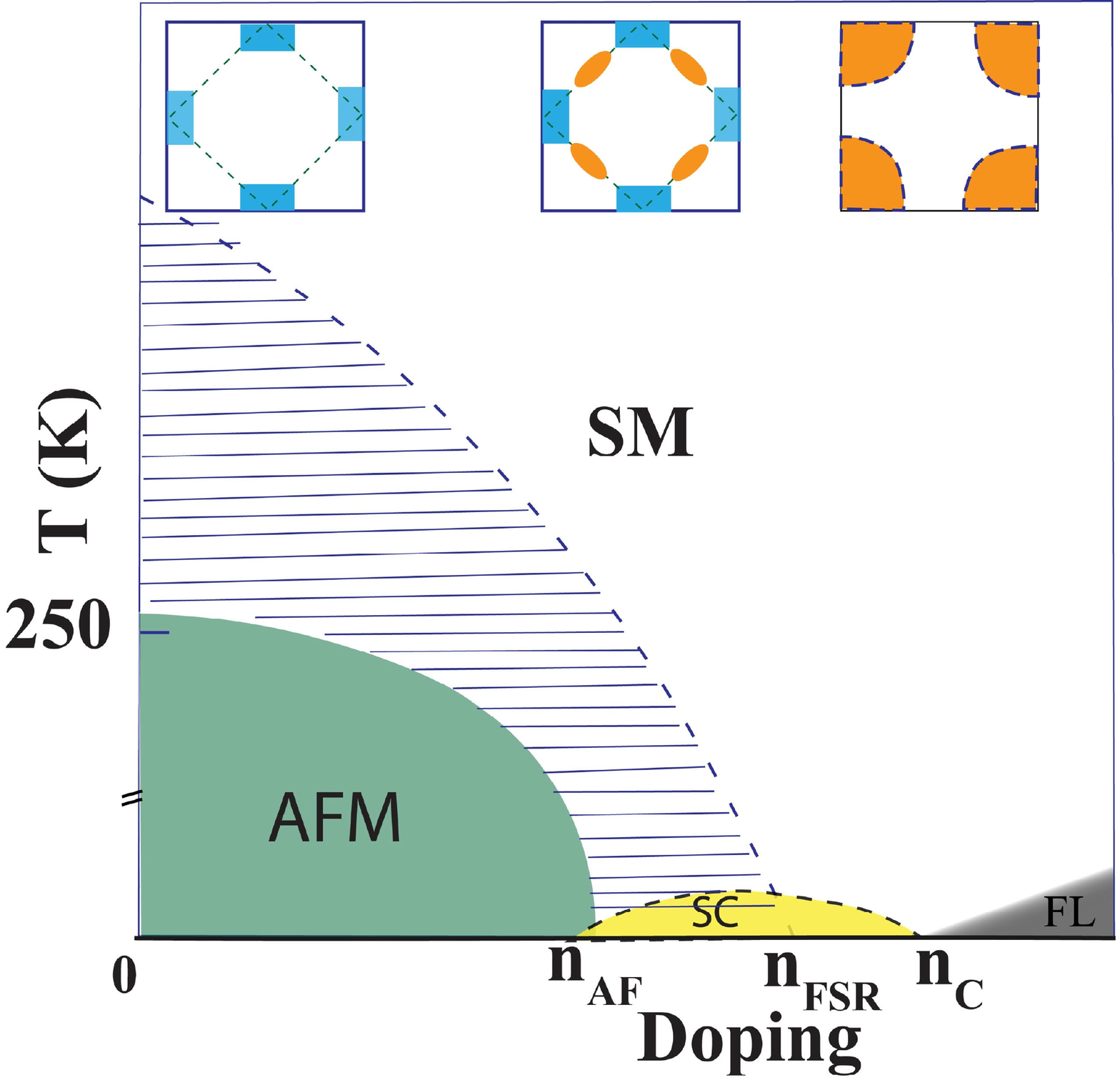}}
\caption{\textbf{Phase diagram of electron doped cuprates (schematic).} Long-range antiferromagnetic (AF) order extends from the Mott insulator state at $n=0$ to $n_{\text{AF}}$, where $n$ is the carrier number. The 2D AF fluctuation region is indicated in blue. The superconducting (SC) region is shown in yellow. The Fermi surface reconstruction (FSR) doping is labelled as $n_{\text{FSR}}$, and the end of the SC region (or ``dome'') is at $n_c$. The white color indicates the strange metallic (SM) region, and Fermi liquid (FL) behavior is found in the black region. The 2D Fermi Surface for various doping regions is shown at the top of the figure. The AF Brillouin zone boundary is indicated by the black dashed lines and the ``hot spots'' at $(\pi,\pi)$ are seen in the middle Fermi surface schematic. The hole regions are orange, and electron regions are blue.}
\end{figure}

Before proceeding, it is worth clarifying what we mean by the phrase ``strange metal.'' The most fundamental distinction between a strange metal and a conventional metal is the absence of well-defined quasi-particles. This is manifested in transport properties which defy conventional theory, the most famous of which is a $T$-linear resistivity that persists from nearly $0$ K to high temperatures above the proposed Mott-Ioffe-Regel (MIR) limit, beyond which Boltzmann theory ceases to be valid. This is in stark contrast to a conventional metal (i.e. a Fermi liquid) in which the low temperature resistivity obeys a $T^2$ power law and the high temperature resistivity saturates when the carrier mean free path is of the order of the lattice constant (or the electron de Broglie wavelength). It has yet to be established whether the high and low temperature behaviors of the strange metal phase are related, so we will take a conservative, experimental point of view and consider them separately. Although there are a number of proposed explanations for the strange metal phase, including the marginal Fermi liquid \cite{varma89}, quantum criticality \cite{qpt, coleman05, sachdev11, berg18, rosch99}, and Planckian dissipation \cite{hartnoll15, zaanen04, zaanensm}, none of these are capable of explaining all experimental results, or are completely accepted. Consequently, in this review we will focus solely on the experimental results and leave their explanation as a theoretical challenge.

\begin{marginnote}[]
\entry{Strange metal}{A poorly understood metallic state which does not conform to conventional theories of transport}
\entry{Mott-Ioffe-Regel (MIR) Limit}{A proposed bound on metallic resistivity motivated by the breakdown of the notion of scattering when the mean free path becomes shorter than the lattice constant of the material}
\end{marginnote}

\subsection{The n-type Phase Diagram}
There are three significant features in the phase diagram of n-doped cuprates: (1) the disappearance of long range AF order at a doping $n_{\text{AF}}$ nearly coincident with the onset of SC, (2) a Fermi surface reconstruction (FSR) at a doping $n_{\text{FSR}}$ caused by a $(\pi,\pi)$ ordering, and (3) the disappearance of SC at a doping $n_c$. These three critical dopings are indicated in Fig. 1 (with $n_{\text{AF}} < n_{\text{FSR}} < n_c$) and the $(\pi,\pi)$ FSR is shown schematically at the top of Fig. 1. The FSR occurs for the wave-vector at which the Fermi surface intersects the 2D AF Brillouin zone boundary, as indicated by the dashed line in the schematic at the top of the Fig. 1. As illustrated in the figure, and verified by experiment, the large hole-like Fermi Surface of the overdoped materials undergoes a FSR to an intermediate region where both hole and electron pockets are present, and then to the underdoped region where only the electron pockets remain \cite{rlgrev}. 

\begin{marginnote}[]
\entry{Fermi Surface Reconstruction (FSR)}{The transformation from a large hole-like Fermi surface to a small Fermi surface with electron and hole pockets as doping is varied}
\end{marginnote}

Charge order is weak and short-ranged in the n-type cuprates \cite{neto15,neto16}, having no apparent impact on their electronic properties, and thus is not shown in Fig. 1. This is in stark contrast to the hole-doped cuprates, where charge order is a significant feature of the phase diagram that competes with the SC \cite{proustannul}. The pseudogap, which has a major impact on the hole-doped phase diagram \cite{statt99,norman05}, is also absent in n-type materials. The onset of 2D AF fluctuations (the dashed blue line in Fig. 1) is commonly referred to as a pseudogap, despite the fact that its physics appears to be unrelated to that of the hole-doped pseudogap. Above the SC dome lies the strange metal phase that is the focus of this review, and beyond the dome is a region where Fermi liquid-like behavior is found. A recent report \cite{tarafm} of ferrmomagnetism observed in this region at temperatures below 4 K will be discussed later.

\begin{marginnote}[]
\entry{Pseudogap}{A mysterious region of the hole-doped phase diagram where the electronic density of states is partially gapped below a crossover temperature $T^\star$}
\end{marginnote}

The transport properties are strongly affected at $n_{\text{FSR}}$ and $n_c$, but not at $n_{\text{AF}}$ \cite{rlgrev,jin11,butch12,taraarx,pampa}. In particular, $n_{\text{FSR}}$ has been determined from dramatic changes in the Hall effect \cite{dagan04}, Angle Resolved Photoemission Spectroscopy (ARPES) \cite{armitage02,matsui07}, quantum oscillations \cite{helm09}, and optical measurements \cite{zimmers}. Meanwhile, $n_{\text{AF}}$ has been determined from inelastic neutron scattering \cite{motoyama07} and low-energy $\mu$SR measrements \cite{saadaoui15}. Recently, a 3D collective charge excitation (distinct from 2D charge order) has been observed in $\mathrm{La}_{2-x}\mathrm{Ce}_2\mathrm{CuO}_4$ (LCCO) thin films with Ce concentration $x= .11$ to $.18$ \cite{hepting18}. This excitation has been attributed to an acoustic plasmon, but its smooth doping dependence suggests that it is not related to any of the principle features ($n_{\text{AF}}$, $n_{\text{FSR}}$, $n_c$) of the n-type phase diagram. The role, if any, of this collective excitation in the SC or normal state properties of the n-type will require future research.

\begin{marginnote}[]
\entry{ARPES}{Angle Resolved Photoemission Spectroscopy}
\entry{Quantum Oscillations}{The periodic modulation of the resistivity in an applied magnetic field which arises due to Landau level quantization of  electronic states, and from which the Fermi surface area may be determined}
\entry{LCCO}{$\mathrm{La}_{2-x}\mathrm{Ce}_2\mathrm{CuO}_4$}
\end{marginnote}

In this review, we discuss the original \cite{takagi89}, most frequently studied, n-type cuprate system: the tetragonal T' phase of $\mathrm{Nd}_{2-x}\mathrm{Ce}_x \mathrm{CuO}_4$ (NCCO), $\mathrm{Pr}_{2-x}\mathrm{Ce}_x \mathrm{CuO}_4$, (PCCO) and LCCO. The preparation of single crystals or c-axis oriented thin films of these materials is complicated by the process of controlling and determining the oxygen content \cite{lambacher, rlgrev}. The carrier doping ($n$) depends on both the $\mathrm{Ce}^{4+}$ concentration ($x$) and the oxygen content, the former of which can be accurately measured while the latter cannot. This has led to confusion in the literature regarding the interpretation of the phase diagram in Fig. 1. There are two reliable methods to determine where a given sample should be located on the phase diagram: the Luttinger count from the Fermi surface area measured via ARPES \cite{wei16}, or the value of the extrapolated $T = 0$ K Hall coefficient ($R_H$). Given that few accurate ARPES studies have been performed on n-type cuprates, we will use $R_H$ ($T \rightarrow$ 0) as our metric for the value of $n$. We note that in the few cases that ARPES and Hall measurements have both been performed on the same materials, the measured values of $n$ agree with one another. 

\begin{marginnote}[]
\entry{NCCO}{$\mathrm{Nd}_{2-x}\mathrm{Ce}_x \mathrm{CuO}_4$}
\entry{PCCO}{$\mathrm{Pr}_{2-x}\mathrm{Ce}_x \mathrm{CuO}_4$}
\end{marginnote}

\begin{figure}
\centering{\includegraphics[width=100mm]{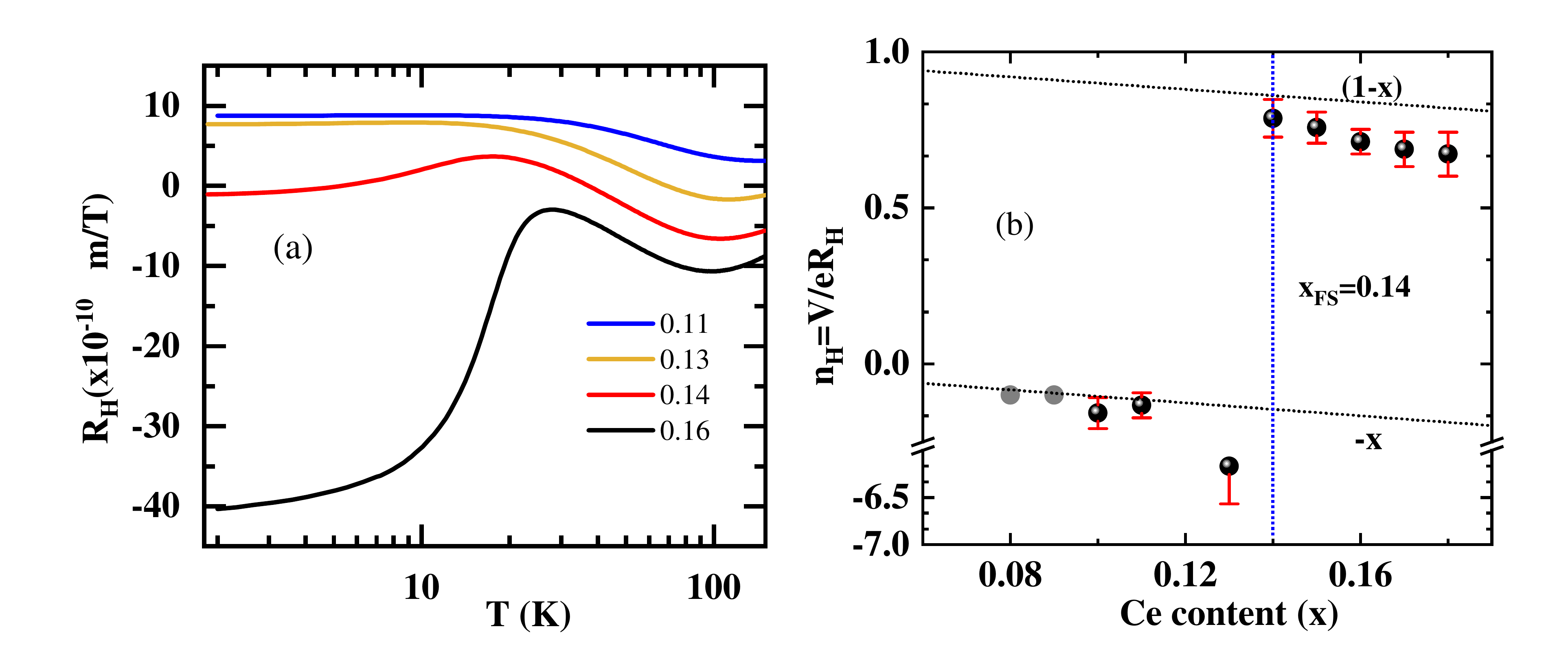}}
\caption{\textbf{Hall effect in La$_{\mathbf{2-x}}$Ce$_\mathbf{x}$CuO$_\mathbf{4}$}. (a) The temperature dependence of the Hall coefficient ($R_H$) at various Ce dopings ($x$) in LCCO near the FSR at $x = 0.14$, (b) The doping dependence of the Hall number $\equiv V/eR_H$ (where $V$ is the unit cell volume) at 2 K. A simple single-carrier doping model would give $n_H = -x$ at low doping and $n_H = 1-x$ for doping above $x_{\text{FSR}}$. A more detailed discussion of the doping dependence of $R_H$ is given in \cite{daganarx}.}
\end{figure}

In Fig. 2a we show Hall data for LCCO thin films.  The dramatic change in the sign and magnitude of $R_H$ at the lowest temperature as a function of $x$ (in these films it was determined that $n \approx x$) is a strong indication of a FSR. In fact, a Hall effect measurement on PCCO films at 350 mK was the first indication of a FSR in an n-type cuprate \cite{dagan04}. The FSR has since been confirmed by quantum oscillations \cite{kartsovnik11}, ARPES \cite{matsui07}, and thermopower \cite{li07} measurements on PCCO and NCCO. For LCCO, the FSR occurs at $n = x = 0.14$, as determined by Hall \cite{tara17}, resistivity \cite{tara17}, and thermopower measurements \cite{pampa}. In Fig. 2b, we plot the Hall number ($V/eR_H$) versus $x$, which dramatically illustrates the FSR at $x = 0.14$ and its impact on the effective carrier concentration above and below the FSR doping. A similar change in Hall number has recently been found in several p-type cuprates at the doping where the pseudogap ends \cite{tailleferrev,proustannul}.

One might expect that the FSR should occur at $n_{\text{AF}}$, where it would be driven by the long-range AF order. However, the most recent ARPES experiments on NCCO \cite{he18} clearly show that the FSR occurs at $n_{\text{FSR}}$, not $n_{\text{AF}}$. Given the existence of short-ranged AF order (with the magnetic correlation length longer than the in-plane lattice constant) in the blue-shaded region of the phase diagram (Fig. 1) which ends at $n_{\text{FSR}}$, it is possible that the FSR is driven by short-range static AF order, as is theoretically expected in a strongly correlated system \cite{senechal02}. Alternatively, topological order could exist between $n_{\text{AF}}$ and $n_{\text{FSR}}$ and cause the $(\pi,\pi)$ band folding \cite{sachdev18}. Although no experimental evidence has been found for such a topological order, it is still a viable explanation, even though short range AF order seems like a more plausible explanation in the n-type cuprates. 

For most reports in the literature -- and all of the results presented here -- one can to good approximation take $n=x$ based on Hall, ARPES, or other data. However, there are a few prior reports where the estimate of $n$, and hence the inferred phase diagram, was incorrect because neither Hall nor ARPES measurements were performed. Notably, thin films of $\mathrm{La}_{2-x}\mathrm{RE}_x \mathrm{CuO}_4$, with $RE$ a $3^+$ rare earth ion (i.e. without Cerium doping) were found to be superconducting with $T_c \sim 25$ K \cite{tsukada05}. It was then claimed, with no supporting transport or ARPES data, that the phase diagram in Fig. 1 was incorrect, and that SC extended down to $n=0$, with no Mott insulating state \cite{matsumoto09,krockenberger13}. Soon after the initial report \cite{tsukada05}, Yu et al. \cite{yu07} completed a thorough transport study on similar T' phase samples with no Cerium doping, and clearly demonstrated that the films were electron-doped, owing to oxygen deficiency. The results of Yu et al. have since been fully verified by Hall, ARPES, and quantum oscillation measurements of non-Cerium-doped films \cite{wei16,horio18,breznay15}, all of which suggest the SC of these films is due to doping from oxygen deficiency, in agreement with the phase diagram shown in Fig. 1. 

\section{TRANSPORT}

\subsection{Overview}

\begin{figure}
\centering{\includegraphics[width=60mm]{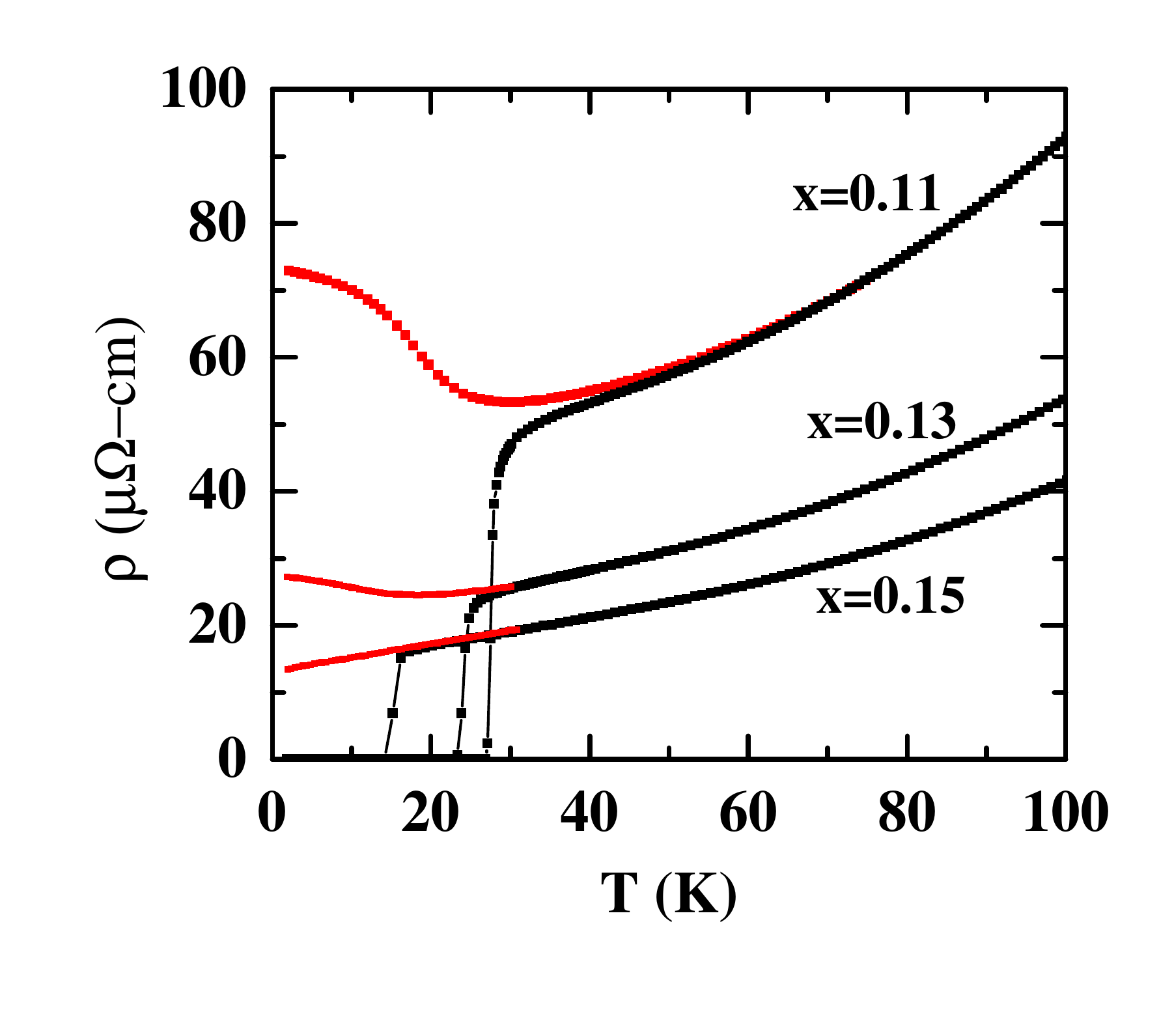}}
\caption{\textbf{Resistivity of La$_{\mathbf{2-x}}$Ce$_\mathbf{x}$CuO$_\mathbf{4}$}. The temperature dependence of the $ab$-plane resistivity at several dopings near $x_{\text{FSR}} = 0.14$. The red curves are the normal state resistivities, measured with $H > H_{c2}$. Below $x_{\text{FSR}}$ a low-temperature resistivity upturn is observed. Above $x_{\text{FSR}}$ the resistivity below $\sim$ 40 K is linear-in-$T$ and above $\sim$ 40 K it is quadratic-in-$T$.}
\end{figure}

Typical n-type $ab$-plane resistivity is shown in Fig. 3 for LCCO with dopings above and below $x_{\text{FSR}} = 0.14$ \cite{tara17}. Above $T_c$ the resistivity follows a power law, $\rho_{ab} \sim T^\alpha$ with $\alpha \sim 2$ up to $400$ K \cite{tara18, daganarx}. Above $400$ K (up to $\sim 1000$ K) the exponent decreases slowly towards $1$, and no resistivity saturation is observed at the estimated Mott-Ioffe-Regal limit \cite{bach11}. This is in stark contrast to conventional metals, and was the first evidence of strange metallic behavior in the n-type cuprates. We will discuss the $T > T_c$ normal state in detail later, but for now we note that the strange metal phase of p-type cuprates is characterized by a strictly linear-in-$T$ resistivity up to temperatures well beyond the nominal MIR limit \cite{martin90, hussey04}, whereas in the strange metallic phase of the n-type materials the resistivity goes as $T^2$. Understanding the strange metallic phase on either side of the phase diagram remains a theoretical challenge, but understanding why the exponent varies between the two families is an even larger mystery. 
 
To probe transport below $T_c$, SC can be suppressed in the n-type materials by applying a transverse (parallel to c-axis) magnetic field of 10 T or less \cite{pampa2}. Typical resistivity data for n-type cuprates is shown in Fig. 3 for LCCO films \cite{tara17} (see \cite{fournier98} for PCCO data). In particular, note the low-temperature resistivity upturn seen in samples with $n < n_{\text{FSR}}$, and the low-temperature linear-in-$T$ resistivity for $x \sim 0.15$ that extrapolates down to $35$ mK. The dramatic changes in the resisistivity and Hall number were interpreted as evidence for an AF quantum phase transition, with a quantum critical point (QCP) at $x \sim 0.165$ for PCCO \cite{dagan04}. Recently, this single QCP interpretation has been called into question by new transport data which suggests that an extended range of low-temperature quantum-critical behavior exists from $n_{\text{FSR}}$ to the end of the SC dome at $n_c$ \cite{jin11,taraarx,pampa}. This is precisely the strange metal phase to be discussed in more detail in the following section. Meanwhile, the low-temperature resistivity upturn seen in underdoped samples below $n_{\text{FSR}}$ and its associated negative magnetoresistance (MR) \cite{dagan05} has been interpreted as arising from spin scattering related to the AF order \cite{hirschfeld, dagan05}. A similar upturn in hole-doped $\mathrm{La}_{2-x}\mathrm{Sr}_x \mathrm{CuO}_4$ (LSCO), the p-type cuprate structurally most similar to the T' phase n-type, is claimed to arise from the loss of carriers associated with the FSR at the end of the pseudogap phase at $p=0.19$ \cite{taillefer2}. A loss of carriers at $n_{\text{FSR}}$ might explain a part of the upturn in the n-type as well, but this will require a more systematic future study.

\begin{marginnote}[]
\entry{Quantum Critical Point (QCP)}{A point lying at zero temperature in a system's phase diagram, across which a quantum phase transition occurs as some parameter (e.g. doping, magnetic field, pressure) is varied}
\entry{MR}{Magnetoresistance}
\end{marginnote}

There have been other proposed explanations for the resistivity and MR of n-type cuprates for $n < n_{\text{FSR}}$ \cite{grevenprl}, particularly for $T > T_c$, but we will not dwell on the underdoped part of the phase diagram, since the resistivity upturn makes any meaningful analysis of the low-temperature transport challenging, and highly sensitive to fitting procedures. Instead, our focus will be the overdoped region where $n > n_{\text{FSR}}$, and strange metallic behavior is most evident. We will primarily discuss data on the LCCO system, as it is the only n-doped material which can be homogeneously doped beyond $n_{\text{FSR}}$, and even beyond $n_c$. By focusing on a particular material, we will change notation and primarily discuss the phase diagram in terms of the Ce concentration $x$ rather than the carrier concentration $n$, with the understanding that for most reports $n \approx x$. To date, it has not been possible to grow single crystals of LCCO of any doping, or of NCCO (PCCO) with $x > .17 (.16) $ \cite{lambacher}. Crystalline, c-axis oriented PCCO films have been prepared with $x > .16$, but not for dopings beyond the end of the SC dome. 

\subsection{Low Temperature Normal State ($T < T_c$)}

\begin{marginnote}[]
\entry{Angular Magnetoresistance (AMR)}{Measurement of the angular dependence of the magnetoreistance as the external field is rotated within the $ab$-plane, which is known to probe spin-charge coupling and is thus a useful tool for investigating magnetic ordering of the copper spins}
\end{marginnote}

\begin{figure}
\centering{\includegraphics[width=120mm]{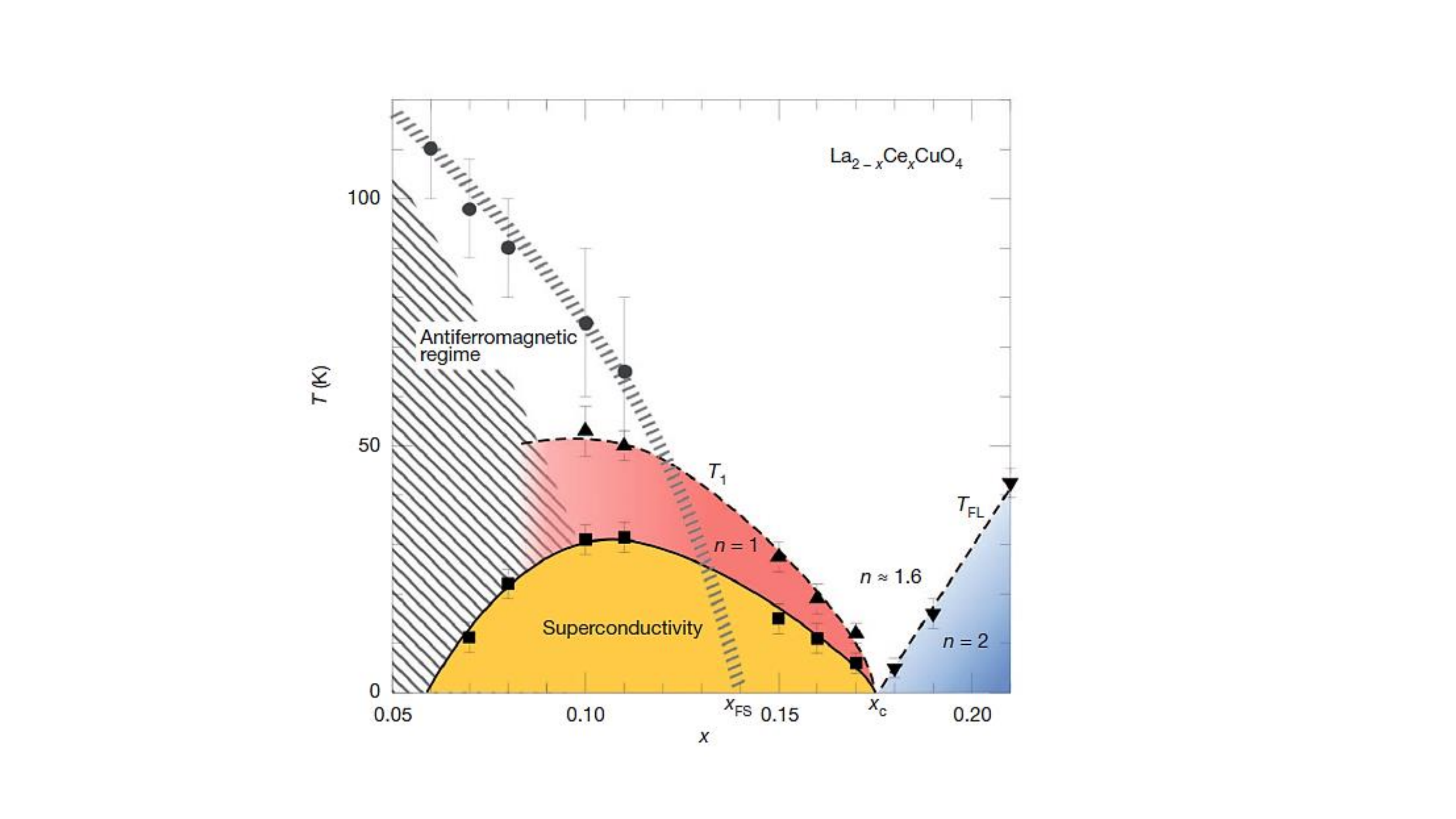}}
\caption{\textbf{Temperature-doping ($\mathbf{T-x}$) phase diagram of La$_{\mathbf{2-x}}$Ce$_\mathbf{x}$CuO$_\mathbf{4}$} (Adapted from \cite{jin11}). In addition to the SC dome (in yellow) and the long-range AF phase (hatched) which ends at $x_{\text{AF}} = 0.08$ \cite{saadaoui15}, the circles indicate the onset of AF fluctuations, as determined by angular magnetoresistance (AMR) experiments \cite{jin09} and the colored regions demarcate the temperature dependence of the resistivity. For all dopings, $\rho \sim T^n$, where $n=1$ in the red region which extends down to mK temperatures when a field is applied to destroy the SC, and $n=2$ in the blue region, where a Fermi liquid-like behavior is seen. Between the two regions, we have the strange metallic phase, extending up to very high temperatures with a different power law. Also note that for this material $x_{\text{FSR}} = 0.14$ and $x_c = 0.175$.}
\end{figure}

The phase diagram for LCCO, based primarily on transport studies, is shown in Fig. 4. For this system $x_{\text{FSR}} = 0.14$, as determined by Hall \cite{tara17}, thermopower \cite{pampa}, resistivity \cite{tara17}, and angular magnetoresistance (AMR) \cite{jin09} measurements. Quantum oscillations from a $x = 0.11$ sample also indicate a small hole pocket, as expected for the reconstructed FS at this doping \cite{josh18}. Quantum oscillations have yet to be observed for the large hole pocket in dopings $x > x_{\text{FSR}}$, or in any other n-type cuprate. The AMR measurement indicates that the FSR is due to a static, short-range, commensurate $(\pi, \pi)$ AF order, which is consistent with the neutron-scattering studies of NCCO \cite{rlgrev,motoyama07}. In most systems with an AF quantum phase transition, a $T$-linear normal state resistivity is found at the QCP, but only at the QCP doping \cite{custers,hayes}. The LCCO system is quite different, having an extended doping range above the putative QCP where a strictly $T$-linear normal state resistivity is observed down to mK temperatures. Some recent data is shown in Fig. 5a. This data is entirely consistent with that reported in \cite{jin11} and shows that $T$-linear normal state resistivity extends down to very low temperatures in the n-type cuprates. This is a manifestation of a very strange metallic ground state that extends over a doping range from $x_{\text{FSR}}$ to $x_c$. Above $\sim 20$ K the resistivity increases above the $T$-linear scattering rate with an approximate $T^2$ behavior from $\sim 60$ K up to beyond $400$ K with no apparent saturation. This higher temperature behavior will be discussed later (see fig 10). Note that for $x > x_c$, the low temperature resistivity follows a conventional $T^2$ behavior \cite{jin11} as shown in Fig. 4.

\begin{figure}
\centering{\includegraphics[width=110mm]{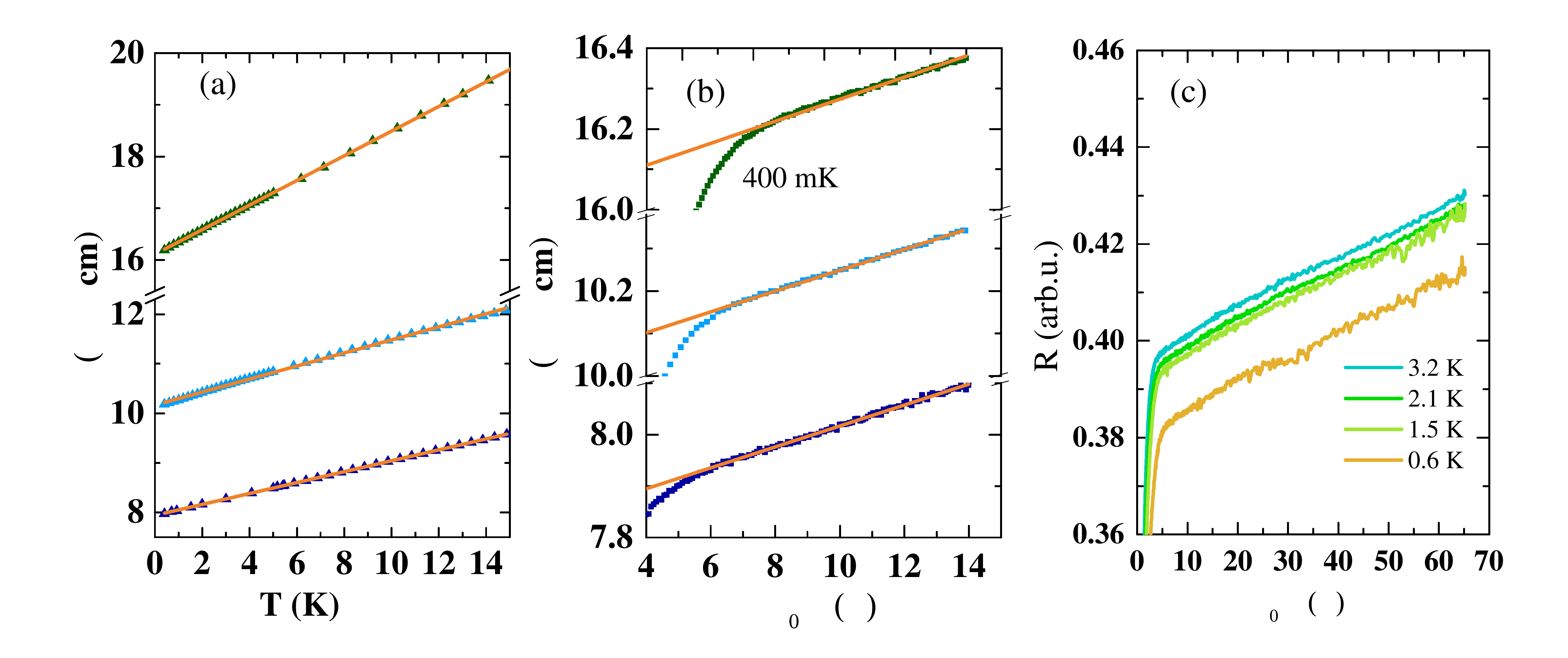}}
\caption{\textbf{Doping-dependent resistivity of LCCO} (adapted from (\cite{taraarx}). (a) $ab$-plane resistivity vs. temperature in the field-driven normal state for LCCO thin films with $x=0.15$ (8 T), $x=0.16$ (7 T), $x=0.17$ (6 T) fitted to $\rho(T) = \rho(0) + A(x)\,T$ (solid orange line), (b) $ab$-plane resistivity vs. magnetic field ($H \parallel \, c$-axis) for $x=0.15, \, 0.16$, and $0.17$ at 400 mK fitted to $\rho(H) = \rho(0) + C(x)\, \mu_0 H$ (solid orange line), (c) Resistivity vs. magnetic field up to 65 T for $x=0.15$ sample at low-temperatures.}
\end{figure}

In the same doping range ($x_{\text{FSR}} < x < x_c$) that strange metallic linear-in-$T$ resistivity is found, an anomalous linear-in-$H$ MR is also observed at low temperatures \cite{taraarx}, illustrated in Fig. 5b at 400 mK (recall that in a conventional metal the MR should go as $H^2$ in low-fields where $\omega_c \tau \ll 1$). Note that this strange metallic MR extends up to 65 T at low temperatures (see Fig. 5c), and crosses over to a conventional low field $H^2$ behavior above $\sim$ 20 K, depending on the doping (see \cite{taraarx} for details).

\begin{marginnote}[]
\entry{Cyclotron frequency}{The frequency $\omega_c = eH/m^\star$ of electron orbits in an applied magnetic field}
\entry{Scattering time, $\tau$}{The average time between electronic scattering events}
\end{marginnote}

This unconventional low-temperature and low-field transport is intrinsic and is not caused by Ce inhomogeneity \cite{rlgrev}. Further, the fact that abrupt changes in properties occur at well defined doping like $x_{\text{FSR}}$ and $x_c$ argues against doping inhomogeneity in the range between these critical dopings. 

\begin{figure}
\centering{\includegraphics[width=60mm]{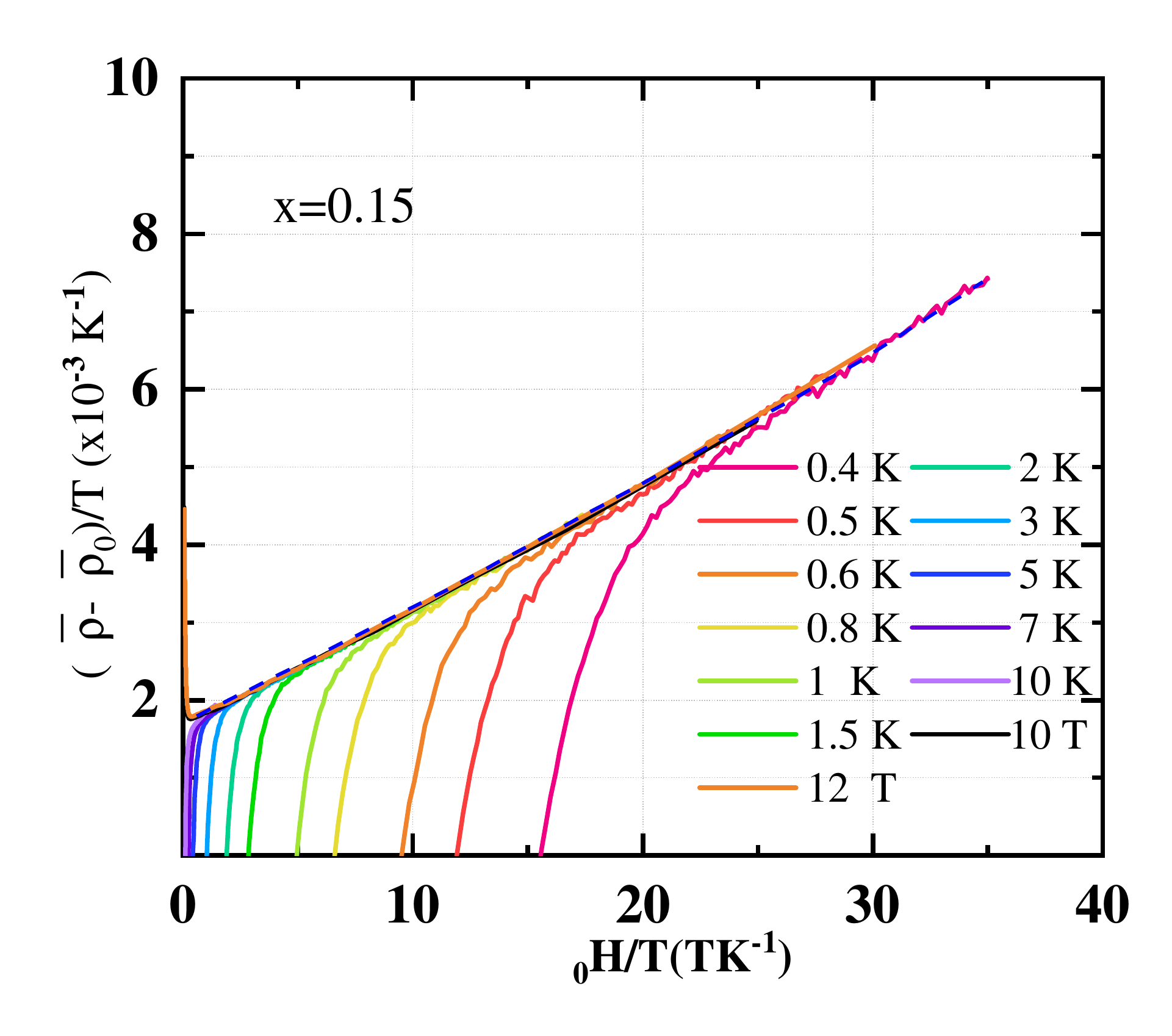}}
\caption{\textbf{Scaling between field and temperature for LCCO with x = 0.15}. (a) Plot of $(\bar{\rho} - \bar{\rho}(0))/T$ vs. $\mu_0 H/T$, where $\bar{\rho} \equiv \rho(T)/\rho(200 K)$ and $\bar{\rho}(0)\equiv \rho(0.4 K)/\rho(200 K)$. This plot has been deduced by varying temperature at fixed field and by varying field at fixed temperature (solid color lines). The curves are fitted to $\Delta \rho = \alpha + \beta (\mu_0 H/T)^\gamma$, with $\gamma = 1.09$ (dashed blue line). See \cite{taraarx} for details.}
\end{figure}

These results are indicative of the scale-invariance (that is, the lack of an intrinsic energy scale) associated with quantum criticality. The MR curves for samples which exhibit linear in $T$ and $H$ behavior all collapse onto a single line when plotted against the energy,

\begin{equation}
\frac{\Delta \rho}{\rho_0} \propto  \varepsilon \equiv A(x) \, k_B T + C(x) \, \mu_B \mu_0 H,
\end{equation}
where $A(x)$ and $C(x)$ are proportional to the slopes of the $T$- and $H$-linear resistivities, respectively \cite{taraarx}. A variant of this scaling analysis is presented in Fig. 6, which suggests that quantum critical fluctuations associated with an extended quantum critical region are responsible for the low-temperature strange metallic behavior in LCCO (and all the T' phase n-type cuprates).

The low-temperature $ab$-plane thermoelectric power (measured in terms of the Seebeck coefficient, $S$) also exhibits strange metallic behavior in the normal state \cite{pampa}, as shown in Figs. 7 and 8. The dramatic change in the temperature dependence of $S/T$ seen in Fig. 7a is another indication of a FSR at $x=0.14$. Above this FSR doping, from $x_{\text{FSR}} < x < x_c$ (the same doping range in which the strange metallic resistivity and MR is observed), $S/T \sim - \ln T$, as seen in Fig. 8. This functional form for $S/T$ is predicted for systems near an AF QCP \cite{paulkot}, which in conjunction with the scale-invariant resistivity and MR furthers the picture of an extended quantum critical region in the LCCO phase diagram. Beyond the SC dome at $x_c$, the low-temperature behavior of $S(T)$ is what one would expect for a conventional metal (where $S \propto T$), as shown in Fig. 7b.

\begin{figure}
\centering{\includegraphics[width=120mm]{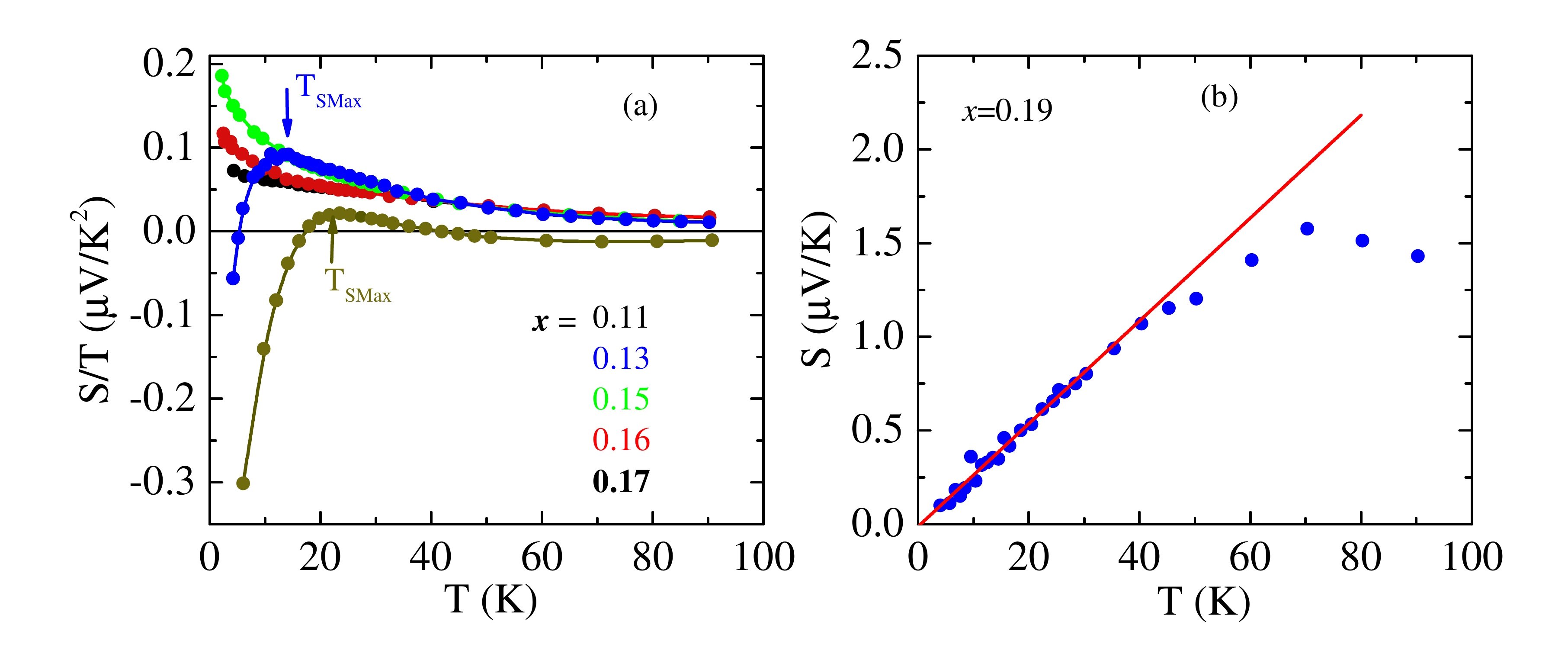}}
\caption{\textbf{Low-temperature normal state Seebeck coefficient (thermopower) of LCCO} (adapted from \cite{pampa}). (a) Seebeck coefficient ($S$) of LCCO for various dopings, plotted as $S/T$ vs. $T$, and measured in an applied magnetic field of 11 T for $x=0.11$ to $0.17$. $T_{\text{S\,max}}$ denotes the temperature below which $S/T$ decreases at low temperatures, reaching negative values for $x=0.11$ and $0.13$. For $x=0.11$ and $0.13$, $S/T$ decreases below 26.5 and 13 K, respectively. For $x=0.15$, $0.16$, and $0.17$ the $S/T$ data increases at low temperature. (b) $S$ vs. $T$ for overdoped LCCO, $x=0.19$ at zero field. The solid line is a fit to $S \propto T$ down to the lowest measured temperature of 4 K.}
\end{figure}

\begin{figure}
\centering{\includegraphics[width=100mm]{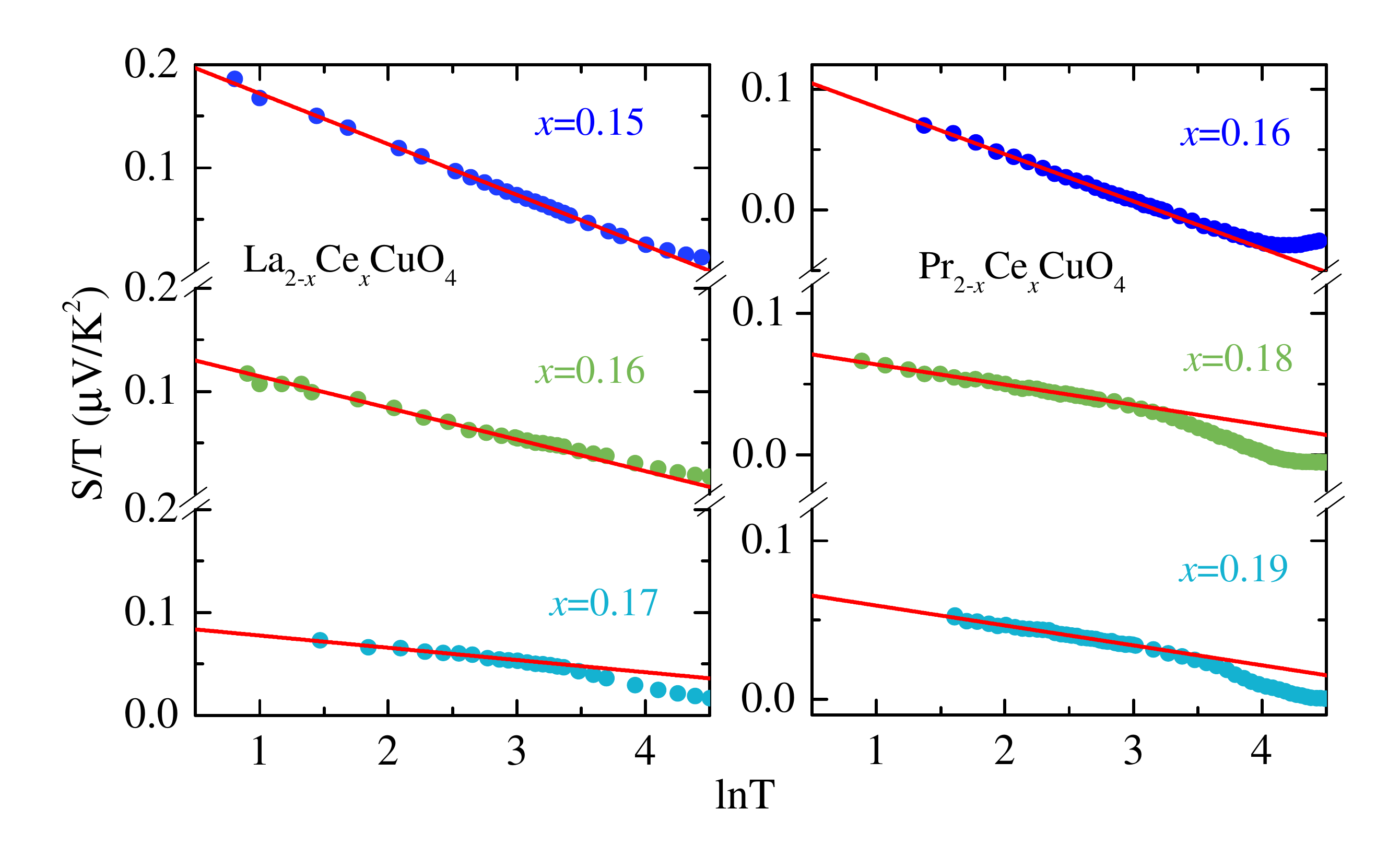}}
\caption{\textbf{Quantum critical temperature dependence of Seebeck coefficient}. Normal state Seebeck coefficient for LCCO films with $x \geq 0.14$ and PCCO films with $x \geq 0.16$, plotted as $S/T$ vs. $\ln T$. The solid lines are a linear fit of the data down to low temperatures. For all samples, $S/T$ exhibits a $-\ln T$ temperature dependence down to the lowest measured temperature of 2 K for LCCO and 3 K for PCCO. A $\ln T$ dependence of $S/T$ has been theorized to result from low energy quasi-two-dimensional spin fluctuations associated with an AF QCP \cite{paulkot}. The magnitude of the $\ln T$ behavior is related to the strength of the coupling between the charge carriers and the spin fluctuations.}
\end{figure}

Further, one can plot the magnitude of $\rho(T), \rho(H)$, and $S/T$ (taken from the slopes of the curves in Figs. 5 and 8) as a function of doping for $x>x_{\text{FSR}}$. This is shown in Fig. 9, together with the $x$ dependence of $T_c$. We see that all of these transport coefficients decrease along with $T_c$ in the overdoped region, which strongly suggests that the origin of the strange metallic behavior is linked to the mechanism of the superconductivity. Although the details of this correlation are unknown, it is important to note that the coefficient of the $\ln T$ thermopower is theoretically linked to the strength of coupling to spin fluctuations, which are claimed to be responsible for the quantum critical behavior \cite{paulkot}. 

\begin{figure}
\centering{\includegraphics[width=150mm]{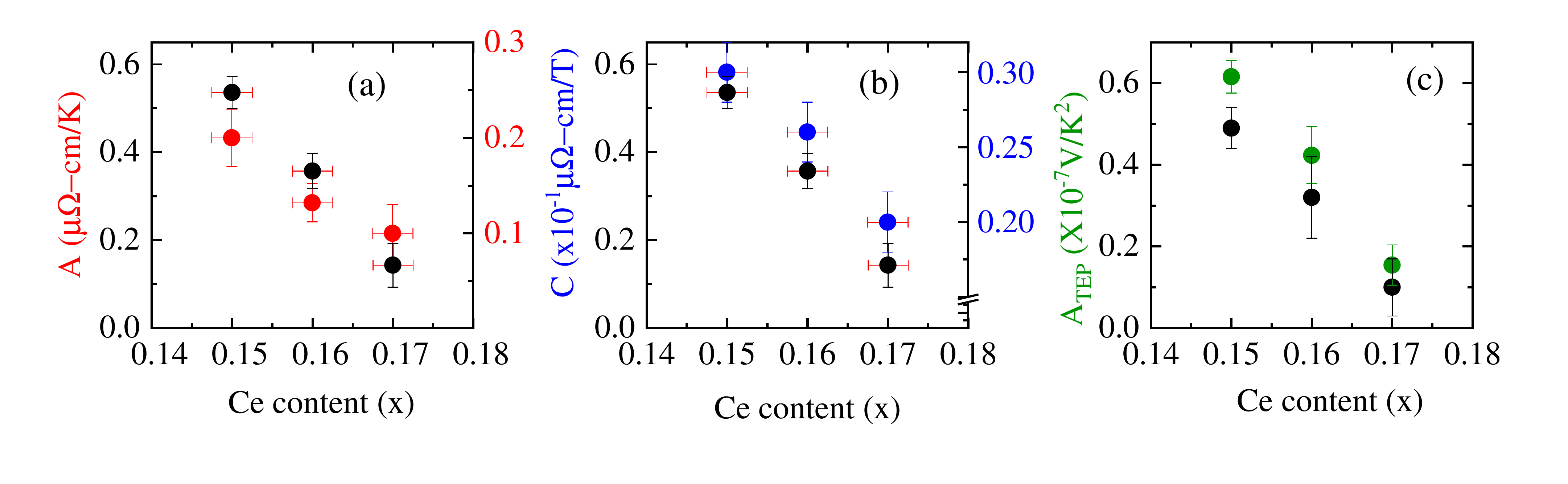}}
\caption{\textbf{Doping dependence of the magnitudes of resistivity, magnetoresistivity, and thermopower for LCCO}. (a) slope ($A$) of linear-in-$T$ resistivity from Fig. 5a (red), (b) slope ($C$) of linear-in-$H$ magnetoresistivity from Fig. 5b (blue), (c) slope ($A_{\text{TEP}}$) of $S/T \equiv A_{\text{TEP}} \ln (1/T)$ from Fig. 8 (green).  The black circles in each plot are $T_c (x)$ normalized to the $T_c$ at optimal doping ($\sim 26 K$). These plots strongly suggest that the origin of the quantum critical resistivity, MR, and thermopower is linked to the cause of the superconductivity. For LCCO (and PCCO) this is most likely spin fluctuations.}
\end{figure}

Very recently, itinerant ferromagnetism has been reported in LCCO doped just beyond the SC dome at temperatures below 4 K \cite{tarafm}.  Unambiguous evidence for static ferromagnetic order in non-SC samples with $x=0.18, 0.19$ has been observed, namely negative $ab$-plane MR and magnetothermopower which both exhibit clear low-field hysteresis, and hysteresis is also seen in the magnetization \cite{tarafm}. None of these features are seen in $x=0.17$ samples, which are inside the SC dome, suggesting the existence of a QCP at $x_c$ between the SC and ferromagnetic phases. Such a QCP would explain the mysterious quantum critical scaling observed near $x_c$ in previous transport studies \cite{butch12}. In fact, the known $\rho \sim T^{1.6}$ behavior of the resistivity near this QCP can now be understood as resulting from ferromagnetic fluctuations about this phase transition \cite{moriya}. Competition between d-wave SC and ferromagnetic order could also answer the fundamental question of why $T_c$ decreases beyond optimal doping, and perhaps the anomalous reduction in superfluid density observed in overdoped cuprates \cite{bozovic16}. Finally, we note that ferromagnetism was conjectured to exist in overdoped p-type cuprates \cite{kopp}, and evidence for ferromagnetic fluctuations have been found in several p-type systems \cite{sonier}. 

\subsection{High-Temperature Normal State}

Above $T_c$, the normal state transport properties of n-type cuprates remain mysterious, and can certainly be called strange metallic. The (zero-field) $ab$-plane resistivity of LCCO for various dopings and $T > T_c$ is shown in Fig. 10 \cite{tara18}, and similar data is found for other n-type cuprates such as PCCO \cite{daganarx} and NCCO \cite{rlgrev,tokura}. From $80$ K to $400$ K (and in some cases, beyond \cite{bach11,tokura}) the resistivity follows a $T^2$ power law for all dopings $x_{\text{AF}} < x < x_c$, which some authors have attributed to conventional Fermi liquid behavior. Of course, in a Fermi liquid $\rho \sim T^2$ only at low temperatures, and such a description is certainly not applicable above room temperature where quadratic temperature dependence persists in the n-type cuprates. The anomalous temperature dependence of the Hall coefficient (see Fig. 2 and \cite{rlgrev}), and consequently the unconventional behavior of $\cot \theta_H$, \cite{dagancot} are in stark contrast to the expected behavior of a Fermi liquid, further undermining the notion that n-type cuprates can be described as Fermi liquids. Further, the optical scattering rate varies as $\omega^1$, as opposed to the $\omega^2$ behavior expected for a Fermi liquid \cite{lobo}. Other weaknesses of this simplistic Fermi liquid interpretation are discussed in detail in Ref. \cite{tara18}. Notably, the $\rho \sim T^2$ behavior of the strange metal phase in n-type cuprates differs from the famous $\rho \sim T$ found in the strange metal phase of p-type cuprates \cite{martin90}. Since $T$-linear resistivity is often found above $\sim 80$ K in conventional metals, arising from electron-phonon scattering (and can be found over a wider temperature range in low carrier density materials such as the cuprates \cite{sankar19}), the high-temperature $T^2$ behavior of the n-type materials is arguably even stranger than the linear-in-$T$ resistivity of the p-type cuprates! Regardless, transport in both classes of materials continues to elude theoretical understanding.

\begin{figure}
\centering{\includegraphics[width=110mm]{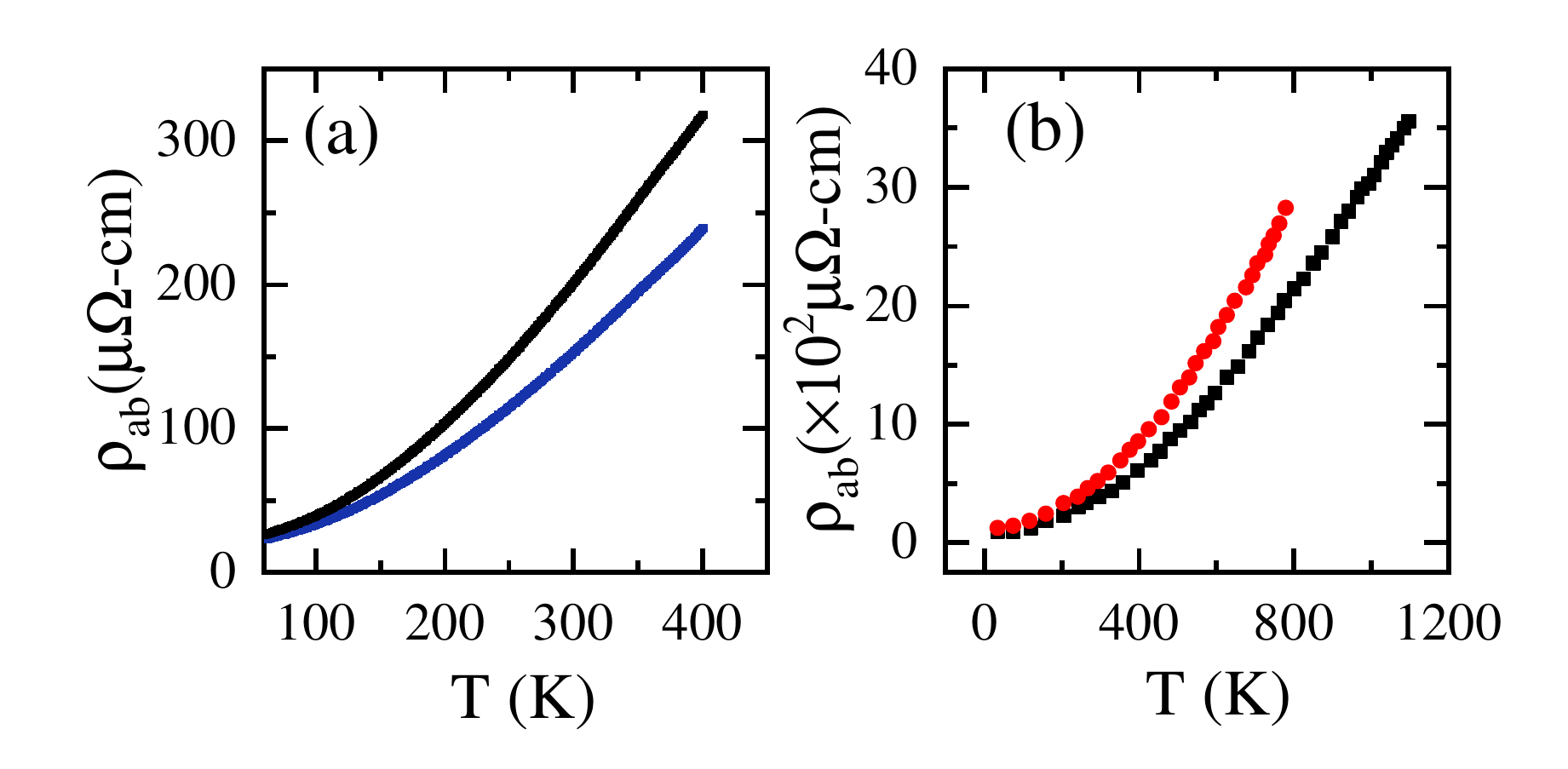}}
\caption{\textbf{High temperature ($\mathbf{T > T_c}$) normal state resistivity}. (a) Resistivity of LCCO for $x=0.15$ (black) and $x=0.17$ (blue). These curves can be fit to $\rho(T) = \rho(0) + A T^\alpha$ with $\alpha = 1.80 \pm 0.02$ (see \cite{tara18}), (b) resistivity of a PCCO film with $x=0.15$ (optimal doping) in red and an NCCO crystal with $x=0.15$ in black (from \cite{bach11}). These curves can be fit to $\rho \sim T^\alpha$ with $\alpha \sim 1.8$ up to $\sim$ 800 K with no sign of resistivity saturation. Some signs of a linear component to the resistivity can be seen above 800 K.}
\end{figure}

The anomalous $T^2$ resistivity persists up to the highest temperatures that it can be measured ($\sim 500-1000$ K, see Fig. 10b), beyond which the films begin to lose oxygen \cite{bach11}), showing no sign of saturation. This suggests that the n-type cuprates violate the MIR limit (this is believed to occur in the p-type cuprates as well \cite{martin90}), although the placement of the bound for a given doping is contentious \cite{tara18}. Yet another manifestation of the strange metallic phase in n-type cuprates is the recently measured $ab$-plane thermal diffusivity of optimally doped crystals from $200-600$ K \cite{thdiff}. The diffusivity was found to vary as $1/T$, with no saturation and a magnitude that could not be explained by phonons alone. To explain this data, it was postulated that the strange metal phase of n-type cuprates can be characterized as an incoherent ``soup'' of strongly interacting electrons and phonons, although future work is necessary to substantiate this picture, as well as to understand why the resistivity varies as $T^2$ but the inverse diffusivity goes as $T$. 

\section{DISCUSSION}
Although this review has focused on the n-type cuprates, it is fruitful to contrast their properties with those of the p-type materials. In particular, the strange metallic state differs considerably between the two sides of the phase diagram: the most striking difference being the temperature dependence of the $T > T_c$ normal state resistivity, which is linear-in-$T$ for the p-type but quadratic-in-$T$ for the n-type (see Fig. 10). However, in both families the power law is robust up to high temperatures (400-1000 K) with no sign of resistivity saturation, leading most of the community to believe that the MIR limit is violated in these systems. Some authors argue that the conventional MIR limit is not appropriate for the cuprates, and is in fact much higher than the typical value of $\sim 150 \mu \Omega$-cm due to either the strong correlations \cite{gunnarson} or low carrier densities \cite{sankar19} typical of cuprate systems, and that consequently the MIR limit is not violated. On the other hand, if the MIR limit is indeed violated above $\sim 300$ K, it is thought that the transport must be incoherent, and that the system must lack the well-defined quasiparticles that facilitate transport in conventional metals \cite{hartnoll15}. The recent thermal diffusivity studies from the Stanford group \cite{thdiff} may prove particularly useful in understanding this picture.

Bold new theoretical proposals have been developed to describe the strange metallic phase (and MIR limit violation) of the cuprates, many of which invoke the notion of ``Planckian dissipation'' \cite{zaanensm, zaanen04}. This is the idea that there is a fundamental bound on the scattering rate $1/\tau$ in any condensed matter system which is saturated by strange metals, where
\begin{equation}
\frac{\hbar}{\tau} \sim k_B T.
\end{equation}
This seems to be a plausible hypothesis for the high-temperature $T$-linear resistivity of the p-type cuprates \cite{martin90}, and has been used to characterize the low-temperature $T$-linear resistivity of n-type cuprates as well as several p-type systems down to 2 K \cite{legros}.  Further, Legros et al. \cite{legros} find a universal value for the slope of this $T$-linear resistivity common to several p- and n-type cuprates, although the origin of this surprising correlation remains unexplained. On the other hand, it is not clear how this idea can be applied to the high-temperature $T^2$ resistivity of the n-type materials, given that the scattering rate exceeds the Planckian bound for $T \gtrsim 25$ K in these materials (i.e. the resistivity exceeds the low-temperature linear-in-$T$ value when extrapolated to higher temperatures). Further, it is puzzling why the low-temperature resistivity would cross over from the Planckian regime to a $T^2$ behavior when the Cerium concentration is just slightly increased from 0.17 to 0.18 in LCCO \cite{jin11}. 

\begin{marginnote}[]
\entry{LSCO}{La$_{2-x}$Sr$_x$CuO$_4$}
\end{marginnote}

Both families of cuprates also undergo FSR's, which can be observed in transport measurements. In the p-type materials the FSR occurs near the end of the pseudogap phase \cite{proustannul}, whereas it occurs as a consequence of short-range AF order in the n-type \cite{rlgrev,he18}. For example, in the LSCO system a FSR occurs at $p_{\text{FSR}} \equiv p^\star$, where the resistivity is linear-in-$T$ down to the lowest measured temperature $\sim 2$ K, in fields up to 80 T \cite{arkady}. However, for $p > p^\star$ up to the end of the SC dome, the resistivity has both $T$ and $T^2$ contributions \cite{cooper09}.  The magnitude of the linear-in-$T$ term decreases as $T_c$ decreases, similar to what is found in the n-type (e.g., Fig. 9). Also, a Fermi liquid-like $T^2$ temperature dependence is observed at low-temperatures in samples doped beyond the SC dome, just as in the n-type materials. It is not clear as to whether the partially linear-in-$T$ resistivity of some p-type cuprates (namely LSCO and Tl2201)  \cite{hussey13} is of the same origin as the pure linear-in-$T$ resistivity of n-type materials such as LCCO (shown in Fig. 5a). One should note that the linear-in-$T$ resistivity in LCCO has been measured down to far lower temperatures (30 mK) than the semi-linear-in-$T$ resistivity of LSCO, which has only been observed down to $\sim 2 K$. Thus, it is possible that the resistivity behavior measured to date in LSCO and Tl2201 may not truly be representative of the ground state of the p-type cuprates. 

Another recently discovered similarity between the p- and n-type cuprates is the mysterious scale-invariant nature of transport (namely, the resistivity as a function of $T$ and $H$) recently reported in LSCO \cite{arkady} and LCCO \cite{taraarx}. In both of these studies, the resistivity and magnetoresistance were found to be simultaneously linear in $T$ and $H$ respectively over certain regions of the doping-temperature-field parameter space. However, the region in parameter space in which this scale invariance was found differs for the two classes of materials. In the p-type it occurs for higher fields and temperatures where $\omega_c \tau \sim 1$, whereas in the n-type it occurs for $\omega_c \tau \ll 1$. Moreover, scale-invariant transport is only found at $p = p^\star$ (where the FSR occurs) in the p-type cuprates, whereas it is found over the entire region $x_{\text{FSR}} < x < x_c$ in the n-type. At comparable dopings, ($p > p^\star$), the p-type systems exhibit a conventional $H^2$ magnetoresistance \cite{rourke11}. Despite this, it is likely that the origin of this scale invariant behavior is the same in both families of cuprates. Considering this scale invariant transport has also been observed in iron-based SC's \cite{hayes} it may be a common feature of not just the cuprates, but high-temperature SC systems in general, and as such is worthy of considerable further investigation. 

\section{CONCLUSION}
We have surveyed the strange metallic normal state of electron-doped cuprates and its signatures in transport measurements, as well as its relationship to the strange metallic phase of the hole-doped materials. The key features of this strange metallic phase are: linear-in-$T$ resistivity and linear-in-$H$ magnetoresistance from $20$ K down to low temperatures ($\sim$ 30 mK) and magnetic fields (up to 65 T); low temperature quantum critical thermopower; and an anomalous range of $\sim T^2$ resistivity from above $T_c$ to well above room temperature (400-1000 K). All of these strange metallic transport behaviors are found over an extended doping range from the FSR to the end of the SC dome. 
Further, there is an surprising correlation between the strength of these strange metallic properties and the superconducting transition temperatures, as shown in Fig. 9. This suggests that the origin of the strange metallic state is intertwined with the origin of high temperature superconductivity in these materials. In the case of the n-type cuprates, it appears likely that quantum critical spin fluctuations play a major role in the physics of both phases. However, none of these strange, non-Fermi-liquid transport properties are theoretically understood. This represents an outstanding challenge for future work, as does better characterizing the relation between the electron- and hole-doped cuprates. After all, realizing the long sought-after goal of a complete theory of the cuprates will undoubtedly involve both sides of the phase diagram, and the differences between the two classes of materials may provide invaluable insights into the mysterious physics of the cuprates.  

\begin{summary}[SUMMARY POINTS]
\begin{enumerate}
	\item The n-type phase diagram is dominated by an AF quantum phase transition at $n_{\text{FSR}}$ where static short-range order vanishes and below which the Fermi surface undergoes a commensurate $(\pi,\pi)$ reconstruction.
	\item There is no psuedogap phase in the n-type cuprates, and charge order (or any other other order) does not have a significant impact on the phase diagram.
	\item The number of doped carriers, $n$, is determined by both oxygen deficiency and $\mathrm{Ce}^{4+}$ doping into the parent compound
	\item The strange metal state of the prototypical n-type cuprate LCCO is characterized by non-Fermi-liquid transport properties, including
	\begin{enumerate}
		\item A linear-in-$T$ resistivity from 30 mK to $\sim$ 20 K (for $x_{\text{FSR}} < x < x_c$),
		\item In the same temperature and doping range, a linear-in-$H$ magnetoresistance for applied fields up to 65 T,
		\item A quantum critical thermopower, $S/T \sim -\ln T$ for $2 K < T < 30 K$ and $x_{\text{FSR}} < x < x_c$,
		\item A robust $T^2$ temperature dependence of the resistivity from $T_c$ to over 400 K, which cannot be explained by Fermi liquid theory.
	\end{enumerate}
	\item There is a strong correlation between $T_c$ and the magnitudes of the $T$-linear resistivity, $H$-linear magnetoresistance, and $-\ln T$ thermopower, suggesting they are all due to quantum critical fluctuations 
	\item For dopings beyond the SC dome, the ground state is a conventional Fermi liquid, which has been found to have ferromagnetic order below 4 K.
\end{enumerate}
\end{summary}

\begin{issues}[FUTURE ISSUES]
\begin{enumerate}
	\item Is the Fermi surface reconstruction driven by short-ranged AF order, or something else (e.g. topological or nematic order)?
	\item Why is there little impact on the transport properties at $n_{\text{AF}}$ where long-range order disappears?
	\item What is the origin of the low-temperature linear-in-$T$ resistivity?
	\item What is the origin of the low-temperature linear-in-$H$ magnetoresistance? 
	\item What causes the apparent quantum critical behavior between $n_{\text{FSR}}$ and the end of the SC dome?
	\item Why does $\rho \sim T^2$ in the $T > T_c$ strange metal phase of the n-type cuprates, but $\rho \sim T$ in the strange metal phase of the p-type?
	\item Is the MIR limit truly violated above 400 K?
	\item Does competition with ferromagnetism explain the decrease of $T_c$ in overdoped n-type cuprates?
\end{enumerate}
\end{issues}

\section*{DISCLOSURE STATEMENT}
The authors are not aware of any affiliations, memberships, funding, or financial holdings that might be perceived as affecting the objectivity of this review. 

\section*{ACKNOWLEDGMENTS}
The authors are appreciative of conversations with Nicholas Butch, Nigel Hussey, Aharon Kapitulnik, Steven Kivelson, Johnpierre Paglione, Sankar Das Sarma, and Louis Taillefer. This work was supported by the NSF Award No. DMR-1708334 and AFOSR Grant No. FA9550-14-10332.

%

\end{document}